\documentclass[aps,prb,twocolumn,superscriptaddress,showpacs]{revtex4-2}
\usepackage{graphicx}  % For including figures
\usepackage{xcolor}
\usepackage{amsmath}   % For advanced math
\usepackage{amssymb}   % For additional math symbols
\usepackage{hyperref}  % For hyperlinks in the document
\usepackage{dcolumn}   % Align table columns on decimal point
\usepackage{bm}        % Bold math symbols

\begin{document}

\title{Interplay between trimer structure and magnetic ground state in Ba$_{5}$Ru$_{3}$O$_{12}$ probed by Neutron and $\mu$SR techniques}

\author{E. Kushwaha}
\affiliation{Rajiv Gandhi Institute of Petroleum Technology, Jais, Amethi, 229304, India}
\author{S. Ghosh}
\affiliation{Rajiv Gandhi Institute of Petroleum Technology, Jais, Amethi, 229304, India}
\author{J. Sannigrahi}
\affiliation{School of Physical Sciences, Indian Institute of Technology Goa, Farmagudi, Goa 403401, India}
\author{G. Roy}
\affiliation{Rajiv Gandhi Institute of Petroleum Technology, Jais, Amethi, 229304, India}
\author{M. Kumar}
\affiliation{Rajiv Gandhi Institute of Petroleum Technology, Jais, Amethi, 229304, India}
\author{S. Cottrell}
\affiliation{ISIS Neutron and Muon Source, STFC, Rutherford Appleton Laboratory, Chilton, Oxon OX11 0QX, United Kingdom}
\author{M. B. Stone}
\affiliation{Neutron Scattering Division, Oak Ridge National Laboratory, Oak Ridge, Tennessee 37831, USA}
\author{Y. Fang}
\affiliation{Jiangsu Laboratory of Advanced Functional Materials, Department of Physics, Changshu Institute of Technology, Changshu 215500, China}
\author{D. T. Adroja}
\affiliation{ISIS Neutron and Muon Source, STFC, Rutherford Appleton Laboratory, Chilton, Oxon OX11 0QX, United Kingdom}
\affiliation{Highly Correlated Matter Research Group, Physics Department, University of Johannesburg, Auckland Park 2006, South Africa}
\author{X. Ke}
\affiliation{Department of Physics and Astronomy, Michigan State University, East Lansing, Michigan 48824, USA}
\author{T. Basu}
\email{tathamay.basu@rgipt.ac.in}
\affiliation{Rajiv Gandhi Institute of Petroleum Technology, Jais, Amethi, 229304, India}

\begin{abstract}
We report a detailed inelastic neutron scattering (INS) and muon spin relaxation ($\mu$SR) investigations of a trimer Ruthenate Ba$_{5}$Ru$_{3}$O$_{12}$ system, which undergoes long-range antiferromagnetic ordering at \textit{$T_N$} = 60 K. The INS reveals two distinct spin-wave excitations below \textit{$T_N$}: one at $\sim$ 5.6 meV and the other at 10-15 meV. By accompanying the INS spectra based on a linear spin wave theory using SpinW software and machine learning force fields (MLFFs), we show that Ba$_{5}$Ru$_{3}$O$_{12}$ exhibits spin frustration due to competing exchange interactions between neighboring and next-neighboring Ru-moments, exchange anisotropy and strong spin-orbit coupling, which yields a non-collinear spin structure, in contrast to other ruthenate trimers in this series. Interestingly, these magnetic excitations do not completely vanish even at high temperatures above \textit{$T_N$}, evidencing short-range magnetic correlations in this trimer system. This is further supported by $\mu$SR spectroscopy, which exhibits a gradual drop in the initial asymmetry around the magnetic phase transition and is further verified through maximum entropy analysis. The results of $\mu$SR spectroscopy indicates a dynamic nature of magnetic order, attributed to local magnetic anisotropy within the trimer as a result of local structural distortion and different hybridization, consistent with canted spin-structure. We predict the ground state of Ru$_{3}$O$_{12}$-isolated trimer through theoretical calculations which agree with the experimentally observed spin excitation.

\end{abstract}

\maketitle

\section{Introduction}
There has been a surge of interest in understanding Ruthenium's unique and versatile magnetic ground states observed in various Ruthenium-based oxide systems, which exhibit a diverse range of physical phenomena, such as,  superconductivity, orbital ordering, quantum spin liquids, metal-insulator transitions, and multiferroicity \cite{ref1,ref2,ref3,ref4,ref5,ref6,ref7}. This unique magnetic ground state arises due to the competing effect of larger 4\textit{d}-orbitals, crystal-electric-field (CEF) effects, and strong spin-orbit coupling (SOC), which can be varied with a small change in crystallographic environments even in the same family. It is predicted that the hybridization of Ru atoms with their neighboring atoms strongly influences the electronic correlations, magnetic interactions, and exchange anisotropy, which give rise to a specific ground state of Ruthenium with distinct physical properties. The unique metal-metal bonding, or Ru(4\textit{d})-Oxygen(2\textit{p}) hybridization resulting from local structural distortions, often leads to fascinating ground states. For example, Ba$_{3}$LnRu$_{2}$O$_{9}$ (where Ln = lanthanide, Y, La, Ce, Nd, Sm, Tb, Ho, Lu), consisting of Ru$_{2}$O$_{9}$ dimers, exhibits versatile magnetic ground states for various Lanthanide ions, even within the same structure \cite{ref7,ref8,ref9,ref10,ref11,ref12,ref13}. A unique spin-3/2 orbital selective Mott ground state of Ru in Ba$_{3}$LaRu$_{2}$O$_{9}$ has been reported, in contrast to the spin-1/2 Ru ground state in Ba$_{3}$YRu$_{2}$O$_{9}$ due to metal-metal bonding, as observed through inelastic neutron scattering (INS) measurements \cite{ref7,ref8}. While Ba$_{3}$CeRu$_{2}$O$_{9}$ features a non-magnetic spin-1 ground state \cite{ref9},Ba$_{3}$HoRu$_{2}$O$_{9}$ exhibits a magnetic ground state with two competing spin structures \cite{ref11}. Recently, a nonmagnetic ground state of a pure RuO$_{2}$ compound is reported through $\mu$SR, spectroscopy \cite{ref14}. The ferromagnetic-metal like system SrRuO$_{3}$ consist of Ru$^{4+}$  spins, while La$_{2}$RuO$_{5}$ shows no-long range ordering containing Ru$^{4+}$ spin configuration due to different crystallographic environment \cite{ref15,ref16}.
In Ru-trimer systems, the well-known compound BaRuO$_{3}$ shows a non-magnetic ground state due to large crystal-electric field (CEF) splitting followed by strong spin-orbit coupling (SOC) producing a J = 0 ground state \cite{ref17}.In an iso-structural compound, Ba$_{4}$Ru$_{3}$O$_{10}$, consisting of corner-sharing Ru$_{3}$O$_{10}$-trimers, the central Ru exhibits a non-magnetic ground state similar to BaRuO$_{3}$, while the two outer Ru atoms form magnetic spin dimers with a gap in the spin excitation spectra \cite{ref1,ref18}.

 \begin{figure}[!t]
{   \centering
    \includegraphics[width=3.5in,angle=0]{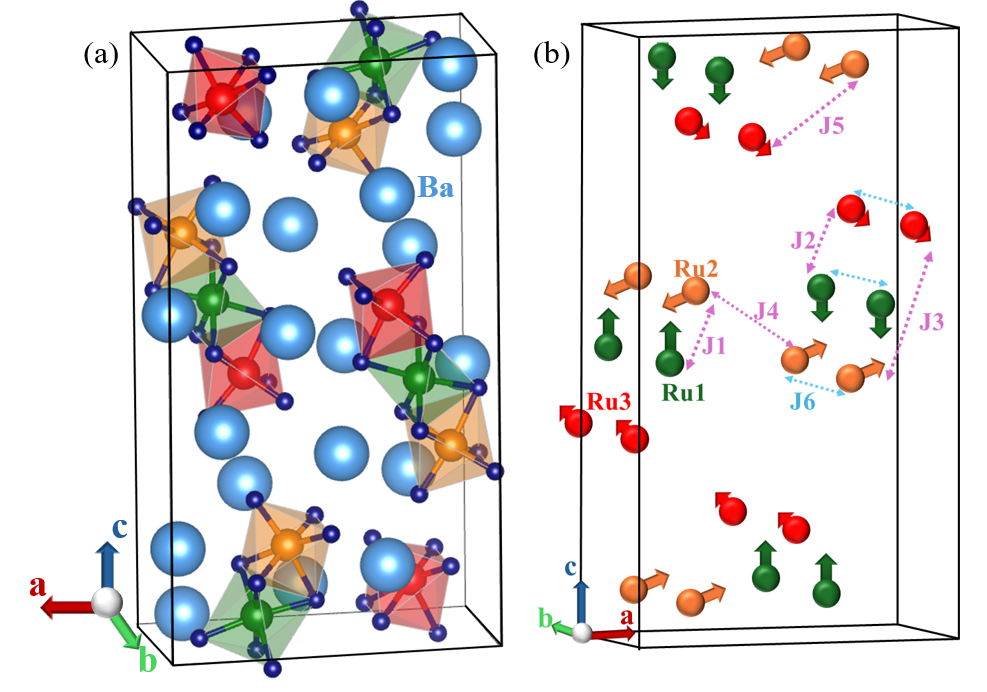}
    \caption{(a) Crystal structure and (b)  Magnetic Structure  of Ba$_{5}$Ru$_{3}$O$_{12}$.}
    \label{figure 1}
}
\end{figure}

In this trimer series, Ba$_{4}$NbRu$_{3}$O$_{12}$, which consists of Ru$_{3}$O$_{12}$ trimers connected through Nb-ions, exhibits long-range magnetic ordering, which yields a S= 1/2 Ru ground state due to strong metal-metal hybridization between Ru-atoms \cite{ref5}. Interestingly, the compound Ba$_{5}$Ru$_{3}$O$_{12}$, which consists of isolated Ru$_{3}$O$_{12}$ trimers (Fig. 1(a)), in sharp contrast to all other trimer compounds in this family, exhibits a long-range magnetic ordering at 60 K without any structural-phase change \cite{ref4}. The neutron diffraction investigation confirms three different nonequivalent Ru-atoms and a non-linear spin-structure, unlike other Ru-trimers in this family( Fig. 1(b)) \cite{ref4}. A reduced moment for ruthenium has been predicted, indicating the possibility of metal-metal bonding \cite{ref4}. The negative Curie-Weiss temperature ($\Theta_{\text{CW}} = -118$ K) indicates the presence of magnetic frustration \cite{ref4}. However, the obtained magnetic moment does not agree with any existing models applicable to other Ruthenates. The nature of the ground state of Ru in this compound remains unclear. However, all other Ruthenate trimers in this family shows non-magnetic ground state or magnetic ground state with collinear spin-structure, the compound Ba$_{5}$Ru$_{3}$O$_{12}$ exhibits a non-collinear spin-structure \cite{ref4}.
  
We have carried out INS, to study the magnetic excitations and the associated exchange interactions present in this compound in detail. We have modelled these INS spectra using SpinW simulations and theoretical calculations based on linear spin wave theory to explore the Ru ground state. Further, we have performed $\mu$SR spectroscopy to investigate the local magnetic field for each Ru-site in this trimer to obtain a clear view of the magnetism present in this compound and local level spin dynamics \cite{ref19,ref20}.

\section{Experimental details}
A polycrystalline sample of Ba$_{5}$Ru$_{3}$O$_{12}$ was synthesized using a solid-state reaction method by mixing high-quality ({$>$}99.9\%) chemicals of BaCO$_{3}$ and RuO$_{2}$, as described in Ref \cite{ref4}. INS experiments were carried out on the fine-resolution Fermi-chopper SEQUOIA spectrometer at the Spallation Neutron Source (SNS) at Oak Ridge National Laboratory (ORNL). Samples were loaded under an atmosphere of helium gas in aluminium cylindrical cans and measured at several temperatures from 4-280 K with incident energy $E_i$ = 30 meV. An empty sample can was measured under identical conditions, and these data have been subtracted from the sample measurements. We have performed the simulation of spin wave using SpinW software \cite{ref19} to understand the intra- and inter-trimer exchange interactions and exchange anisotropy in the title compound. We have analyzed INS data using Mantid software which used standard Mantid scripts to convert our neutron scattering measurements to standard histograms in energy and detector space\cite{ref20}.  We have used DAVE software for generating scattering intensity and normalized scattering intensity as a function of momentum transfer, $|$Q$|$, and energy transfer\cite{ref21}. Calculations were performed employing the INSPIRED software \cite{ref22} using pre-trained machine learning force fields (MLFFs) based upon the MatterSim modelling package to calculate the optimized structure and calculate the momentum dependent phonon modes \cite{ref23}.

We have done the zero-field (ZF) $\mu$SR measurements using the EMU spectrometer at the ISIS neutron and muon facility to determine the local level spin dynamics. In a $\mu$SR experiment, implanted positive muons interact with the local internal field at the muon site in the sample. After 2.2 $\mu s$, muons decay into one positron and two neutrinos. We detect the positrons, which are preferentially emitted in the direction of the muon spin at the time of decay. By that, we can trace the polarization of the muon-spin ensemble \cite{ref24}. The muon asymmetry was calculated through the counts measured in the forward and backward detectors placed with respect to the initial muon-spin polarization direction $N_{F, B}$ and corrected using a parameter $\alpha$, which accounts for the relative efficiency and geometrical asymmetries between the forward and backward detectors., via $A(t) = (N_{F} - \alpha N_{B})/(N_{F} + \alpha N_{B})$. The calculated value of $\alpha$ is 0.84. The asymmetry is directly proportional to the polarization of the muon ensemble. For the analysis of $\mu$SR data, we have used Mantid \cite{ref20} and WiMDA software \cite{ref25}. 

\section{Results and Discussions}
\subsubsection{Inelastic Neutron Scattering: Spin-wave excitations and Short-Range spin-correlation}
The background-subtracted scattering intensity S ($|$Q$|$,E) of Ba$_{5}$Ru$_{3}$O$_{12}$ as a function of energy transfer (E) versus momentum transfer ($|$Q$ |$) is shown in Fig. 2 for selective temperatures of T=4, 30, 100, 200, and 280 K. A strong intense feature is observed around 10-15 meV in low-$|$Q$|$ region (0.5 $<$$|$Q$|$$<$ 1.5) below the magnetic ordering temperature(see Fig. 2(a-b) for 4 K and 30 K respectively). This intense feature vanished at higher temperatures in the paramagnetic region (see Fig. 2(d-e) for 200 K and 280 K, respectively). A careful observation suggests the presence of a weak excitation around 5.6 meV at 4 K and 30 K which get suppressed at high temperatures in the paramagnetic region. These excitations are further confirmed in the one-dimensional energy cut obtained by integrating a fixed  $|$Q$|$ -region ranging from 0.5 to 1.5 Å$^{-1}$ (intensity versus energy transfer plot in Fig. 2(f)). Fig. 2(f) exhibits a broad, intense peak from 10-15 meV and a small peak around 5.6 meV at 4 K and 30 K. Both of these features become suppressed or highly damped at higher temperatures. The excitations in the low-$|$Q$|$ region are considered to be of magnetic origin because the magnetic form factor of the scattering intensity decreases with increasing $|$Q$|$, whereas the phonon excitation is usually observed at higher $|$Q$|$-regions \cite{ref26}. The absence of these features above the magnetic ordering temperature is also consistent with these features being associated with the magnetic moments. Hence, we characterize this feature as a spin-wave excitation. 
Interestingly, we observe a broad weak excitation at 100 K in a large energy range from 6-16 meV at low-$|$Q$|$ region (see Fig. 2(c)), which is documented in the form of a broad low-intense peak in I vs E plot in Fig. 2(f). Such a broad and weak feature above T$_N$ often arises from diffuse scattering which is attributable to short-range magnetic ordering from the Ru-trimers. The negative Curie-Weiss temperature (-118 K) \cite{ref4} is consistent with the presence of short-range spin correlation at 100 K.

\begin{figure}[!t] 
{   \centering
    \includegraphics[width=3.5in,angle=0]{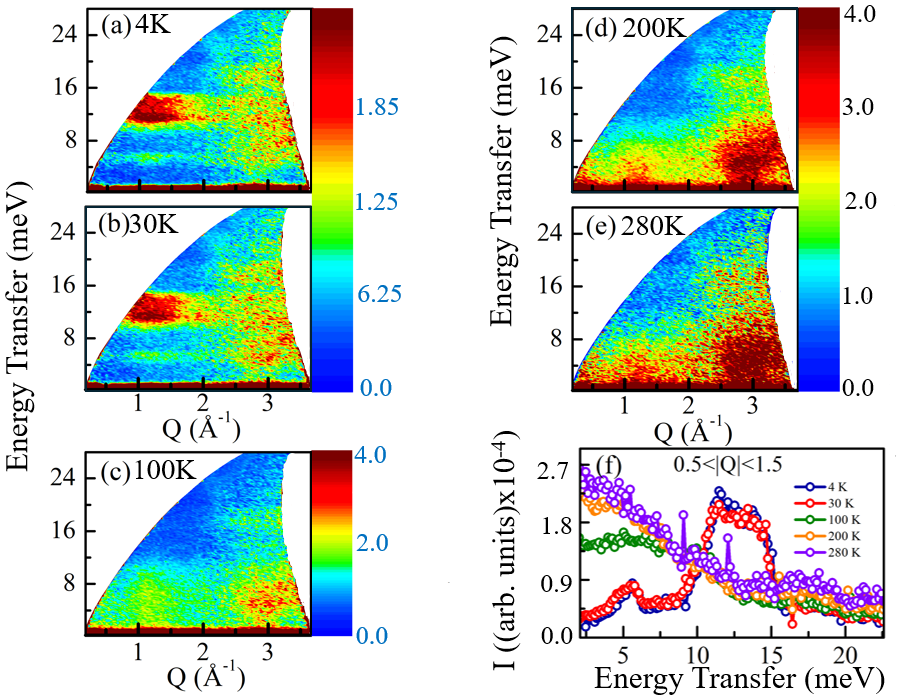}
    \caption{ Color-coded contour maps of the measured scattering intensity as a function of energy and momentum transfer ($|$Q$|$) of Ba$_{5}$Ru$_{3}$O$_{12}$ using 30 meV incident energy neutrons (a) at 4 K, (b) at 30 K, (c) at 100 K, (d) at 200 K, (e) 280 K and (f) The $|$Q$|$- integrated, Intensity vs energy transfer for temperatures between 4 K and 280 K.}
    \label{Figure 2}
}
\end{figure}

To further confirm these features as being spin-excitations, we examine the Bose Factor normalized scattering intensity. This quantity is proportional to the dynamic susceptibility $\chi"$($|$Q$|$,E). We first extract the background subtracted (Can subtracted) and Bose factor normalized spectrum at each individual temperature. We then subtract the treated data (Can subtracted and Bose factor normalized) at T = 280 K from the lower temperature data, which is referred as Bose factor correction. The result of this for T= 4 K and T= 100 K is shown in Fig. 3(a) and (b).This treatment of the data illustrates the well localized bands of magnetic scattering in the spectrum between 10 and 15 meV energy transfer and around 5.6 meV energy transfer. The difference in the dynamic susceptibility also shows how these features are damped and significantly broadened but are still present even at T= 100 K. The enlarged view of 5.6 meV feature is shown in Fig. 3(c) for the 4 K data. These excitations are observed for 30 K with a slightly reduced intensity, as expected for magnetic systems with increasing temperature. The broad weak feature in INS spectra at 100 K (Fig. 2(c)) still persists even after Bose factor correction (Fig. 3(d)), confirming the persistence of magnetic fluctuations for a range of temperatures greater than T$_N$. This spectral weight is not significant for T= 200 K as shown in the scattering intensity data of Fig. 2(d) as well as the Bose factor corrected data shown in Fig. 3(d). For temperatures greater than 100 K, it is likely that the magnetic fluctuations are within a paramagnetic regime and would be contributing to a broad diffuse scattering at and near zero energy transfer.
\begin{figure}[!t] 
{   \centering
    \includegraphics[width=3.5 in,angle=0]{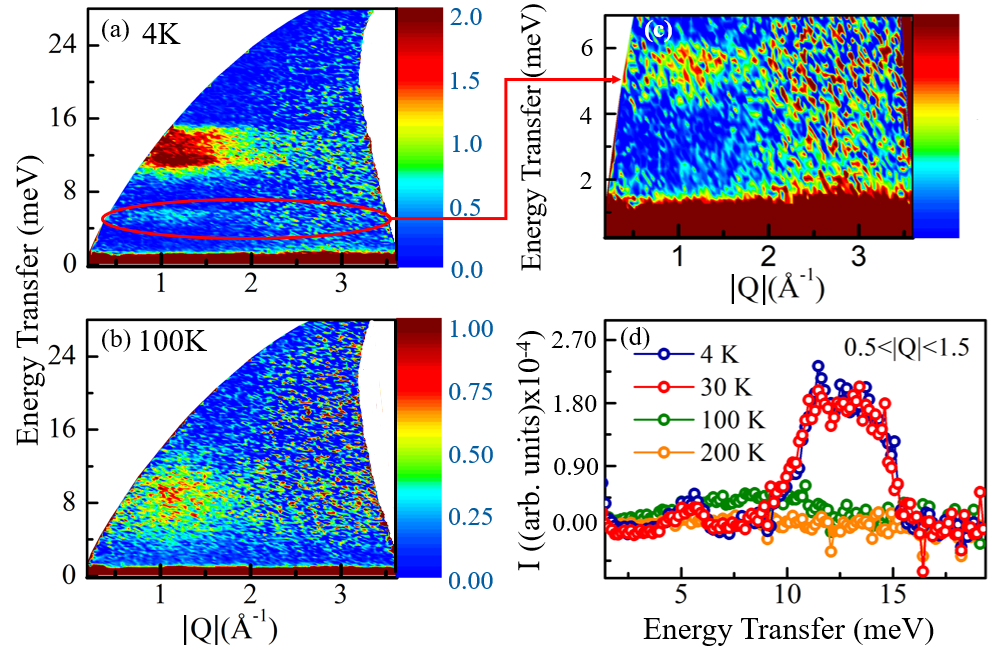}
    \caption{Bose factor corrected data (as described in text) of scattering intensity as a function of momentum transfer and energy transfer for Ba$_5$Ru$_3$O$_{12}$.The data is first Can subtracted and Bose factor normalized then subtracted from T= 280 K treated data from each data set shown in figure. Panels (a)-(c) are shown on the same color scale as indicated. (a) T= 4 K, (b) T= 100 K , (c) enlarged view of T= 4 K data with intensity scaled by factor of 2, (d) Bose factor corrected scattering intensity as a function of energy transfer, measured over the temperature range 4–200 K. The data are integrated over the momentum range 0.5$<$$|$Q$|$$<$1.5 Å$^{-1}$. }
    \label{Figure 3}
}
\end{figure}

This compound has three inequivalent Ru-sites with different magnetic moments. This fact was confirmed in our earlier neutron diffraction results (see Ref.\cite{ref4} ). The 5.6 meV excitation is prominent around $|$Q$|$$\sim 1.1~\text{\AA}^{-1}$ which corresponds to the magnetic Bragg peak (010), which was observed in prior 10 K neutron powder diffraction results (see Ref.\cite{ref4} and Supplementary S.I.Fig. 1 \cite{ref27}). We have checked that the intensity of (010) in neutron diffraction is only influenced by the magnetic moment of Ru1 atom, which is in the middle of the trimer (see S.I.Fig. 1 in S.I \cite{ref27}). Hence, we conclude that this 5.6 meV spin-excitation is related to the Ru1-spin. The weak intensity of this spin-excitation might be due to the contribution of weak effective exchange interactions originating from various competing exchange interactions, J1 (Ru1-Ru2) and J2 (Ru1-Ru3). The broad, intense 10-15 meV peak could result from a combination of multiple peaks, governed by the combined dominant exchange interactions involving the Ru1, Ru2, and Ru3 magnetic atoms.

\subsubsection{Exchange-interaction and ground state calculations}
To determine the relevant exchange interactions, we have performed SpinW simulations to compare with the experimental INS spectra \cite{ref19}. Neutron diffraction results suggest an anisotropy along the c-axis, where all the spins tend to align having Ru1, Ru2, and Ru3 ordered moments of 1.52 $\mu_B$, 1.36 $\mu_B$, and 0.91 $\mu_B$ respectively\cite{ref4}. The Ru1 moments aligned exactly along the c-axis, while Ru2, and Ru3 moments are slightly canted with the c-axis in the ac-plane due to exchange frustration. The spin-Hamiltonian for this system, considering all these factors, is expressed as:

\begin{equation}\label{eq1}
\ H  = \sum_{i<j} \vec{S}_i^{\,\mathrm{T}} \mathbf{J}_{ij} \vec{S}_j + \sum_i \vec{S}_i^{\,\mathrm{T}} \mathbf{D} \vec{S}_i
\end{equation}
Where
\[
\vec{S}_i = 
\begin{pmatrix}
S_i^x \\
S_i^y \\
S_i^z
\end{pmatrix}
\quad , \quad
\mathbf{J}_{ij} = 
\begin{pmatrix}
J_{ij}^{xx} & J_{ij}^{xy} & J_{ij}^{xz} \\
J_{ij}^{yx} & J_{ij}^{yy} & J_{ij}^{yz} \\
J_{ij}^{zx} & J_{ij}^{zy} & J_{ij}^{zz}
\end{pmatrix}
\quad \text{and} \quad 
\]
\\
\[
\mathbf{D} =
\begin{pmatrix}
D^{xx} & D^{xy} & D^{xz} \\
D^{yx} & D^{yy} & D^{yz} \\
D^{zx} & D^{zy} & D^{zz}
\end{pmatrix}
\]
In the Hamiltonian \( \mathbf{J}_{ij} \) is the $3 \times 3$ anisotropic exchange tensor, and \( \mathbf{D} \) is the single-ion anisotropy tensor. All these interactions occur between two nearest neighbor Ru atoms. The exchange interactions are shown in Fig. 1(b) and also tabulated in Table 1.The matrix representation of each exchange interaction \(\mathbf{J}_{ij} \) is provided in the Supplementary Information~\cite{ref27}. The anisotropy is uniaxial, with a non-zero component along the \( z \)-direction, \( \mathbf{D}_{zz} = -2.1 \)~meV, while the components along the other directions are zero. 
The solution of the Hamiltonian with the simulated $J$-values reproduces a spin structure that closely agrees with the experimentally obtained magnetic structure from neutron diffraction. The simulated spin-wave excitations from this Hamiltonian exactly mimic the experimental INS spectra, yielding a strong spin excitation around 10–15 meV and a weak spin excitation around 5.6 meV, shown in Fig. 4(a) (2D contour plot of intensity vs $|$Q$|$ ). The one-dimensional energy cut in (intensity vs. energy transfer) exactly replicates the two excitations with experimentally obtained features in Fig. 4(b).

\begin{figure}[!t] 
{   \centering
    \includegraphics[width=3.5in,angle=0]{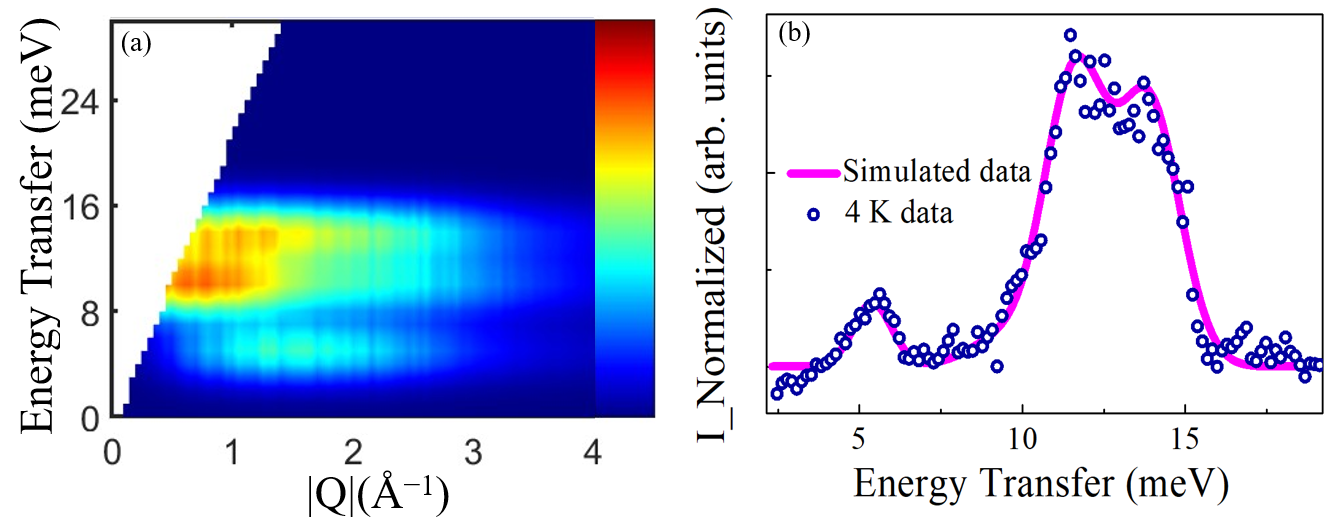}
    \caption{(a) Simulated color-coded contour map of the calculated spin-wave excitation spectrum as described in text as a function of momentum transfer \( |Q| \) and energy transfer and (b) Intensity vs Energy transfer where line shows the simulated data Obtained from SpinW and open circles show the experimental data at 4 K for $|$Q$|$= 0.5 to 1.5 Å$^{-1}$.The solid line in (b) has been linearly scaled to match the background and scattering intensity of the measurement.}
    \label{Figure 4}
}
\end{figure}
\
We have checked all the possibilities by tuning the different parameters in the spinW simulations. The exchange anisotropy and J$_4$, J$_5$, J$_6$ is required to exactly replicate the experimentally obtained non-collinear spin-structure along with the intra trimer exchange interactions. Only by tuning intra-trimer exchange interactions (J$_1$, J$_2$, J$_3$), we did not obtain the desired spin-wave excitations, that is, one weak excitation at 5.6 meV and another broad intense excitation between 10-15 meV as observed in the INS experiment. However, 30\% variations in J$_4$-J$_6$ do not significantly change the excitation energies, but do affect the stability of the magnetic structure which is obtained from experimental neutron diffraction data. The extracted exchange parameters from spinW simulations are summarized in Table 1, which reveals the dominant antiferromagnetic (AFM) interactions within the trimer units. The Ru1-Ru2 bond length ($\sim$2.5~\AA) is shorter than the typical metallic Ru-Ru distance ($\sim$2.65~\AA), suggesting enhanced orbital overlap and a dominant direct AFM exchange interaction ($J_{1} = 4.2$~meV) between Ru1 and Ru2. The comparatively smaller value of the effective exchange interaction between Ru1 and Ru3 ($\sim$2.7~\AA, $J_{2} = 1.8$~meV) could result from an average value of competing direct exchange interaction and superexchange interaction. The significant AFM exchange interaction between Ru2 and Ru3 ($J_{3} = 4.5$~meV)  competes with both $J_{1}$ and $J_{2}$, which gives rise to a non-collinear AFM configuration in this trimer system. The large value of $J_{3}$ is attributed to a superexchange mechanism, likely due to favorable orbital overlap and bond angles that enhance hopping integrals via intervening anions. Such unconventional behavior where next-nearest-neighbor exchange interactions are stronger than effective nearest-neighbor interactions have been reported in several other complex systems with strong spin-orbit coupling~\cite{ref28,ref29,ref30,ref31,ref32}. We endorse a similar reason where the interplay between strong spin-orbit coupling and different degrees of hybridization for particular Ru-sites within the Ru-trimers is responsible for different competing exchange interactions. The weak but non-negligible inter-trimer exchange interactions (see Table 1) further support the development of long-range magnetic order. Overall, our findings emphasize that, in such correlated systems, competing exchange interactions, magnetic anisotropy, and spin-orbit coupling collectively play a decisive role in stabilizing non-collinear magnetic structures. 
 
\begin{table}[htb]
\caption{\label{Table1} Intra-trimer and inter-trimer Anisotropic exchange interactions obtained from SpinW (negative value for FM and positive value for AFM). The distances are as depicted in Fig.~1(b).}
\begin{ruledtabular}
\begin{tabular}{lll}
Label & Component & Value (meV) \\
\hline
\( J_1 \) (intra) & \( J_{zz} \) & 4.2 \\
\( J_2 \) (intra) & \( J_{xz} \) & 1.8 \\
\( J_3 \) (intra) & \( J_{zx} \) & 4.5 \\
\( J_4 \) (inter) & \( J_{zx} \) & 0.3 \\
\( J_5 \) (inter) & \( J_{zx} \) & 0.1 \\
\( J_6 \) (inter) & \( J_{zx} \) & -0.06 \\
\end{tabular}
\end{ruledtabular}
\end{table}

Further, we have theoretically calculated the spin state and excitation energy due by considering only dominant exchange interactions ($J_1$-$J_3$) observed in the SpinW models. While the uniaxial anisotropy term $D_{zz}$, having a magnitude comparable to $J_1$–$J_3$, was included in the full SpinW simulations to capture the detailed excitation spectrum, it has been excluded from this analytical model. The goal of presenting this Hamiltonian is to offer a qualitative understanding of the spin dynamics without the added complexity of anisotropic interactions. A detailed theoretical DFT study may provide further insight into this complex system. A simplified form of the Hamiltonian (in equation 1) for the Ru$_{3}$O$_{12}$-trimer is expressed below:
\begin{equation}\label{eq2}
\ H =  J_{1} (\vec{S}_1 \cdot \vec{S}_2) + J_{2} (\vec{S}_1 \cdot \vec{S}_3) + J_{3} (\vec{S}_2 \cdot \vec{S}_3)
\end{equation}
\\
The eigenvalues corresponding to this trimer system are given by \cite{ref33}:
\begin{equation}\label{eq3}
\begin{split}
E(S_{12}, S) = & \frac{J_1}{2} \big[S_{12}(S_{12} + 1) - S_2(S_2 + 1) - S_1(S_1 + 1)\big] \\
               & + \frac{J_2}{2} \big[S(S + 1) - S_{12}(S_{12} + 1) - S_3(S_3 + 1)\big] \\
                & + \frac{J_3}{2} \big[S(S + 1) - S_{2}(S_{2} + 1) - S_3(S_3 + 1)\big]
\end{split}
\end{equation}
\\
\\
Here, $S_1$, $S_2$, and $S_3$ denote the spin operators for Ru1, Ru2, and Ru3, respectively. The spin moments are $\frac{3}{2}$ for Ru1 and Ru2, and 1 for Ru3. Ru1 is located in the middle of the trimer.

To fully characterize the trimer states, we introduce additional quantum numbers, $S_{12}$ and $S$, which are derived from the vector sums:
\[
S_{12} = S_1 + S_2 \quad \text{and}\quad S = S_1 + S_2 + S_3,
\]
with constraints:
\[
0 \leq S_{12} \leq 2S_1 \quad \text{and} \quad |S_{12} - S_3| \leq S \leq S_{12} + S_3
\]

The trimer states are defined by the wave functions $\lvert S_{12}, S \rangle$, and their degeneracy is given by $(2S + 1)$. By solving this equation and substituting the $J$ values from SpinW, we obtain different eigenstates. The detailed calculations for determining the energy eigen values are shown in the Supplementary Information (S.I.) \cite{ref27}. 
\begin{figure}[!t] 
{   \centering
    \includegraphics[width=3.5in,angle=0]{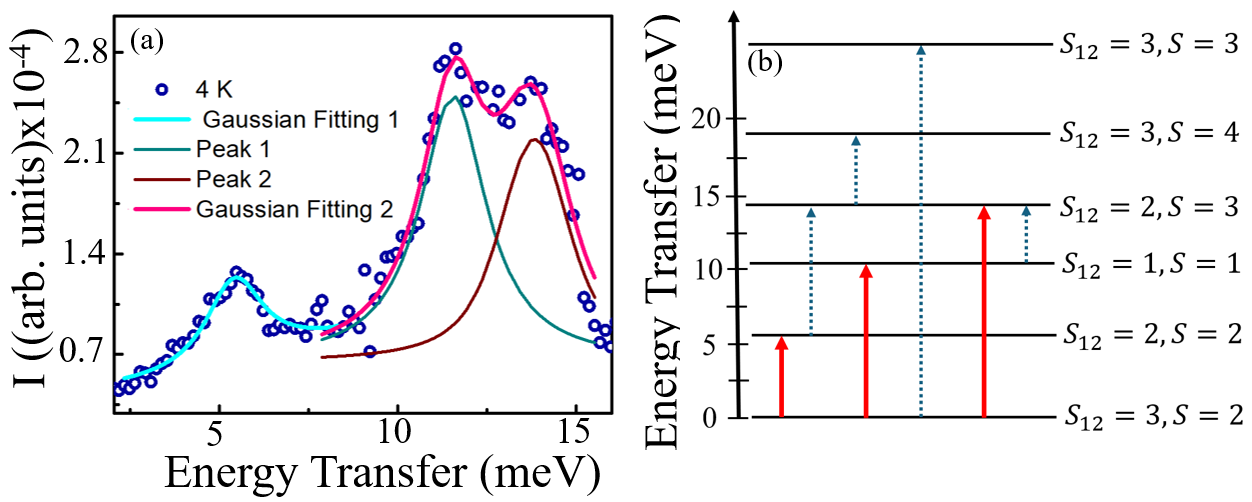}
    \caption{(a) Gaussian fitting of 4 K Intensity vs Energy Transfer plot, (b) Theoretically predicted low-lying energy levels with an $S=2$ ground state. Here, $S_{12}$ indicates the coupling between $S_1$ and $S_2$, and $S$ represents the total spin. We observed the three red-arrow transitions. A blue-arrow transition may exist, but we could not detect them at 60 K(ground state transition) as well as at high temperatures (excited state transitions).}
    \label{Fig.5}
}
\end{figure}
We have fitted the experimental INS spectra using three Gaussian peaks, one Gaussian for the ~ 5.6 meV excitation and two Gaussians for the broad excitation between 10-15 meV, as shown in Fig. 5(a). We assign these excitations as transitions from the ground state to excited states at $\varepsilon_{1}$,$\varepsilon_{2}$ and $\varepsilon_{3}$.
From the calculations, we find that the transition from  \( |3,2\rangle \) states to  \( |2,2\rangle \), \( |1,1\rangle \) and \( |2,3\rangle \) corresponds to an excitation energy of  5.85 meV, 10.5 meV and 14.7 meV respectively by following the selection rule \( \Delta S = 0, \pm 1 \), which is consistent with the observed excitations in the INS spectra depicted in Fig. 5(a). The energy level diagram for this trimer state is shown in Fig. 5(b) \cite{ref34}. Hence, the trimer ground state should be characterized as \( |3,2\rangle \). In \( |3,2\rangle \), S=2 is the ground state of Ru-trimers, which represents the total spin moment of the Ru$_3$O$_{12}$ trimers. Using this, we have calculated the magnetic moment \( 4\, \mu_B \) and the effective magnetic moment to be \( 4.89 \, \mu_B \),  which agrees with experimentally obtained magnetic moment from Neutron diffraction results and effective moment from magnetic susceptibility \cite{ref4}. Therefore, we conclude that the observed spin-excitation is restricted to Ru$_3$O$_{12}$ -trimers.

\subsubsection{Origin of excitation at high-$|$Q$|$ region}
Our earlier discussion primarily focused on the low-$|$Q$|$ region, a closer inspection of the high-$|$Q$|$ region in Fig. 2(a-e) reveals intriguing features. Notably, excitations at a similar energy range as the low-$|$Q$|$ magnetic scattering are clearly visible even at high-$|$Q$|$. To explore this, we plotted the intensity vs $|$Q$|$ across the full-$|$Q$|$ range for two selected energy regions around 5.6 meV and between 10-15 meV shown in Fig. 6(a) and (b). At low $|$Q$|$ ($<$1.5 Å$^{-1}$), the intensity decreases with increasing $|$Q$|$, which is consistent with magnetic excitations, as the magnetic form factor reduces with increasing $|$Q$|$.  However, at higher $|$Q$|$, the scattering intensity increases proportional to Q$^2$, consistent with a phonon cross-section. This suggests that phonons begin to dominate for $|$Q$|$$>$ 2 Å$^{-1}$.\\

To clarify the role of phonons, we have performed phonon calculations based on the published crystal structure \cite{ref4}. The calculated phonon modes shown in S.I.Fig. 2 span up to approximately 22 THz (approximately 91 meV) with no indication of negative phonon energies.  We have also used this calculated spectrum in the INSPIRED software to calculate the powder averaged neutron scattering intensity convolved with the instrumental energy and momentum resolution.  This calculation also included multiple phonon processes and temperature dependent thermal population of phonons. We present these results in S.I.Fig. 3 for both scattering intensity and Bose factor normalized scattering intensity for T=4 K and T=280 K.  These calculations provide a good characterization of the extent of the observed phonon scattering in the measurement.  The range of momentum transfers used to characterize magnetic scattering, 0.5$<$ $|$Q$|$$ <$ 1.5 Å$^{-1}$, is shown to have a small phonon scattering contribution at elevated temperatures.  Likewise, these calculations provide additional clarity regarding the relative scattering intensity and Bose factor normalized intensity of the phonon contribution as a function of momentum transfer in the range of larger momentum transfer, 2$<$ $|$Q$|$$ <$ 3.5 Å$^{-1}$. There exists calculated phonons in the higher momentum transfer region that overlap significantly with the magnetic excitations we have characterized at smaller momentum transfer.\\

Fig. 6(c) shows energy cuts at high $|$Q$|$ (2$<$ $|$Q$|$$ <$ 3.5) across various temperatures, showing a prominent peak in the vicinity of 5.6 meV, a similar energy to the low-$|$Q$|$ lower energy spin-excitation. However, the temperature dependent scattering intensity is difficult to interpret. As seen in Fig. 6(c) the peak at larger momentum transfer softens as a function of temperature from 6.2 meV at 4K to 4.9 meV at 280 K. Such behavior could be attributed to phonon softening above the magnetic ordering temperature due to electron-phonon coupling, which might be influenced by changes in the electronic correlations around magnetic transition, or due to magnon-phonon coupling, as reported in many other systems, such as, YMnO$_{3}$, Sr$_{14}$Cu$_{24}$O$_{41}$ and CaMn$_7$O$_{12}$ compounds  \cite{ref36,ref37,ref38}. This phonon softening can be parameterized by the relation:
\begin{equation}\label{eq3}
\omega_{\text{phonon}} = \omega_0 + \lambda S(Q ,T)
\end{equation}

where $\omega_{\text{phonon}}$ is the modified phonon frequency due to spin-phonon coupling, $\omega_{\text{0}}$ is the uncoupled phonon frequency,  $\lambda$  is the spin-phonon coupling constant, and S(Q,T) is the spin correlation function, which depends on temperature and momentum. However, such a shift could also be a consequence of the change in spectrum due to thermal population effects over the large range of temperatures examined. To consider this, we have plotted the Bose factor normalized data in Fig. 6(d). We find that the spectral weight as shown in Fig. 6(d) does not shift significantly with temperature, excluding the possibility of magnon-phonon coupling being observed from these measurements.\\

To gain further insight, we have plotted the Bose factor corrected intensity for the momentum transfer range of 0.5$<$ $|$Q$|$$ <$ 1.5 Å$^{-1}$ in S.I.Fig. 4(a) and 2$<$ $|$Q$|$$ <$ 3.5 Å$^{-1}$ in S.I.Fig. 4(b). This treatment of the data allows one to observe the proposed magnetic excitations around 10-15 meV for T= 4 K and 30 K to be observed even in what would typically be considered a high-$|$Q$|$ region of the spectrum (S.I.Fig. 4(b)).The low-energy (5.6 meV) excitation is weaker in intensity but clearly visible. The presence of magnetic scattering in this range of larger values of momentum transfer is likely due to the extended range of the Ru$^{4+}$ magnetic form-factor.
In order to investigate whether the high-$|$Q$|$ excitations arise solely from phonons or include a magnetic contribution, we calculated the magnetic form factor for Ru as a function of momentum transfer using standard analytical expressions for 4\textit{d} transition metals (see S.I.Fig. 4(c)) \cite{ref35}. By comparing the intensity ratios between low and high $|$Q$|$ regions in the Bose factor corrected data (See S.I.Fig. 4(a) and (b)), we find that at $|Q| \approx 3.5$~\AA$^{-1}$, only about $\sim 3.9\%$ of the magnetic form factor remains. To better understand this behavior, we have taken the average value of $|$F(Q)$|$$^2$  for low  $|$Q$|$ (0.5-1.5 Å$^{-1}$ ) and high $|$Q$|$ (2$<$ $|$Q$|$$ <$ 3.5) range, which is 0.7652 and 0.1944 respectively. Using this , we have estimated the expected magnetic intensity at high $|$Q$|$:
\begin{equation}
I_{\mathrm{high}\,Q} = I_{\mathrm{low}\,Q} \left( \frac{ |F(Q)_{\mathrm{high}_{\mathrm{avg}}}|^2 }{ |F(Q)_{\mathrm{low}_{\mathrm{avg}}}|^2 } \right)
\end{equation}

Substituting the observed values (from Fig.~3(d) and S.I.Fig. 4(a)):

\begin{align}
I_{\mathrm{high}\,Q} &= 2.32 \times 10^{-4} \times \frac{0.1944}{0.7652} \nonumber \\
&\approx 0.59 \times 10^{-4}
\end{align}
\begin{figure}[!t] 
{   \centering
    \includegraphics[width=3in,angle=0]{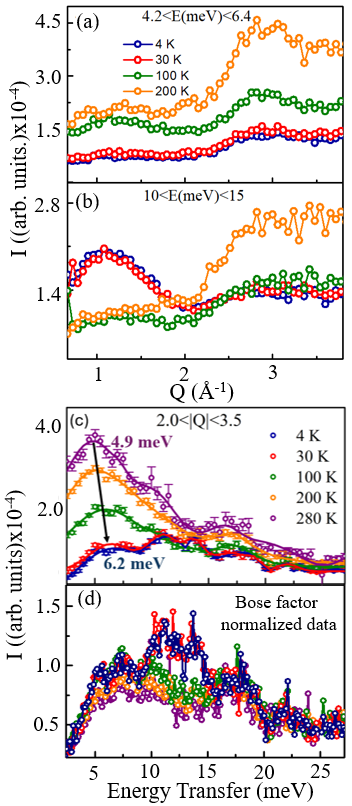}
    \caption{Intensity vs momentum transfer at Fixed Energy (a) E=10 - 15 meV (b) 4.2 - 6.4 meV and The $|$Q$|$- integrated, Intensity vs momentum transfer of (c) Raw data at 4 K - 280 K that shows the 5.6 meV peak shift with temperature, (d) Bose factor normalized data at 4 K – 280 K} 
    \label{Fig.6}
}
\end{figure}
\
This value nearly agrees with the experimental intensity of 0.67 x 10$^{-4}$ observed for 10-15 meV excitation in S.I. Fig. 4(b) after Bose factor correction. This result supports that the high-$|$Q$|$ feature at 10–15 meV is of magnetic origin, and likely due to the range of the magnetic form factor of the Ru ion.\\ 
We do note that our phonon calculations indicate the presence of a large phonon density of states in the vicinity of the higher energy magnetic mode. A thorough single crystal INS measurement along with additional phonon calculations to understand which modes involve the Ru-sites in the crystal structure would help to quantify any signature of spin-phonon coupling.
\subsubsection{Muon Spin Relaxation}
 Fig. 7(a-e) represents ZF-$\mu$SR spectra of Ba$_{5}$Ru$_{3}$O$_{12}$ \cite{ref39}. The initial asymmetry drops gradually without developing muon spin oscillation due to the high magnetic damping in the antiferromagnetically ordered phase as the temperature goes below T$_{N}$ = 60 K \cite{ref40, ref41}. We missed the oscillation as it falls within the detector dead time of the ISIS pulsed source. Two distinct relaxations are clearly visible in the asymmetry spectra. The ZF-$\mu$SR spectra are fitted with the sum of a simple exponential decay function and an exponential decay oscillation function, with the addition of a flat background constant term.
\begin{equation}\label{eq5}
G_{Z} (t) = A_{1} exp(-\lambda_{1} t) + A_{2} cos(\omega t + \phi) exp(-\lambda_{2} t) + A_{bg}
\end{equation}
Here, $A_{1}$ and $A_{2}$ are the initial asymmetry parameters, $\lambda_{1}$ and $\lambda_{2}$ are the muon relaxation rates, $\omega$ is the muon precession frequency, $\phi$ is the phase of initial oscillation, and $A_{bg}$ is the temperature-independent flat background term. The relative efficiency of the detectors, $\alpha$, was calculated in diamagnetic mode using the transverse field (TF) data at 20 G and base temperature (TF 20). We then used this $\alpha$ = 0.84 value consistently across all data points. Next, we estimated the background by fitting a flat background function to the data at the lowest, highest, and intermediate temperatures. Background determination was carried out using both Mantid and WiMDA software. In both cases, we obtained background values of approximately 0.15 and 15, respectively. The value of 15 in WiMDA corresponds to its use of percentage asymmetry in plots. For all temperature data, we fixed the $A_{bg}$ at 0.15, which usually corresponds to the muons stopped inside the silver holder or might get planted inside the sample, where the internal field distribution was very negligible \cite{ref40,ref42}. The oscillatory and non-oscillatory term in the fit function corresponds to two different muon sites. In the paramagnetic phase, at the higher time domain (above 5 $\mu s$), the decoupling effect is visible with the parallel shifting of the fitting curves, as shown in Fig. 7(f).
\begin{figure}[!t] 
{   \centering
    \includegraphics[width=3.5in,angle=0]{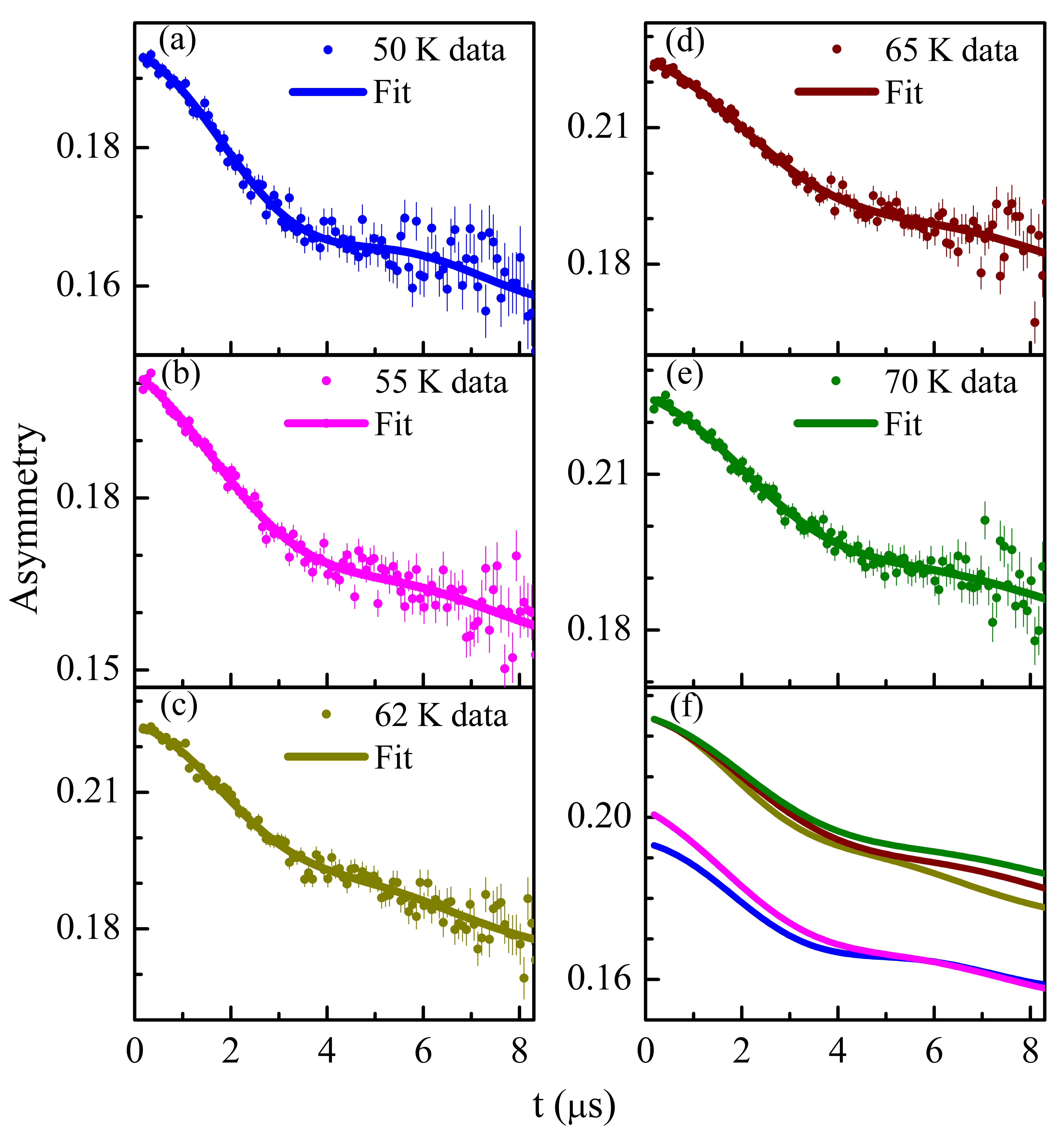}
    \caption{ (a) - (e) Zero-field Asymmetry data between 50 K to 70 K is shown. We have fitted the data with the addition of two exponential decay functions, with a temperature-independent flat background term. Solid circles, which are the experimental data, and solid lines are the fittings. (f) A combined plot of only fitting curves for all temperatures is shown.}
    \label{Fig.7}
}
\end{figure}

\begin{figure}[!t] 
{   \centering
    \includegraphics[width=3in,angle=0]{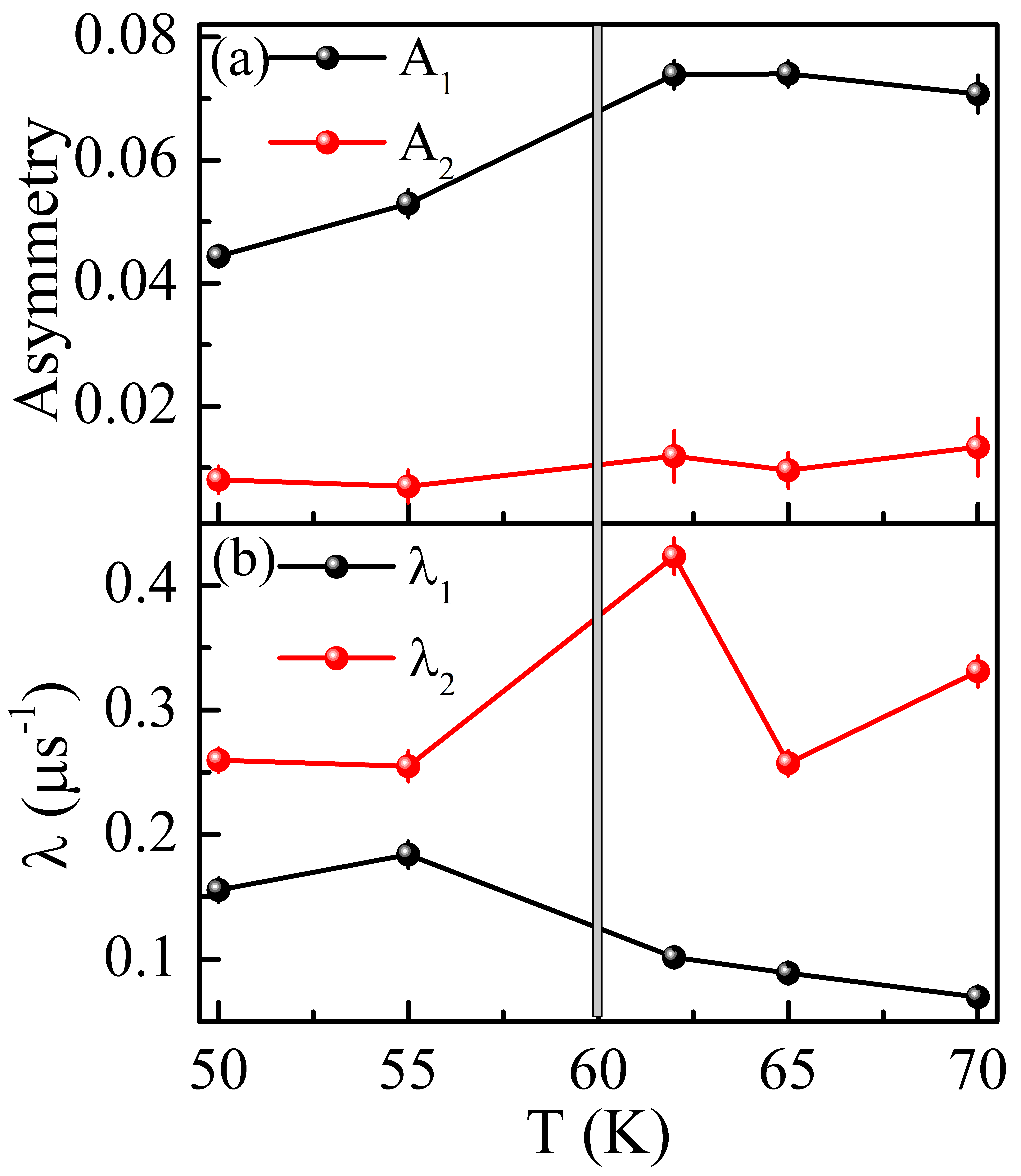}
    \caption{ (a) Initial asymmetry parameters $A_{1}$ and $A_{2}$  and (b) the muon relaxation rates $\lambda_{1}$ and $\lambda_{2}$ as a function of temperature are shown. $A_{1}$ and $\lambda_{1}$ correspond to the first exponential decay function, and $A_{2}$ and $\lambda_{2}$ correspond to the second exponential decay function in the fit function. Solid lines are only a guide to the eye.}
    \label{Fig.8}
}
\end{figure}

\begin{figure}[!t] 
{   \centering
    \includegraphics[width=3in,angle=0]{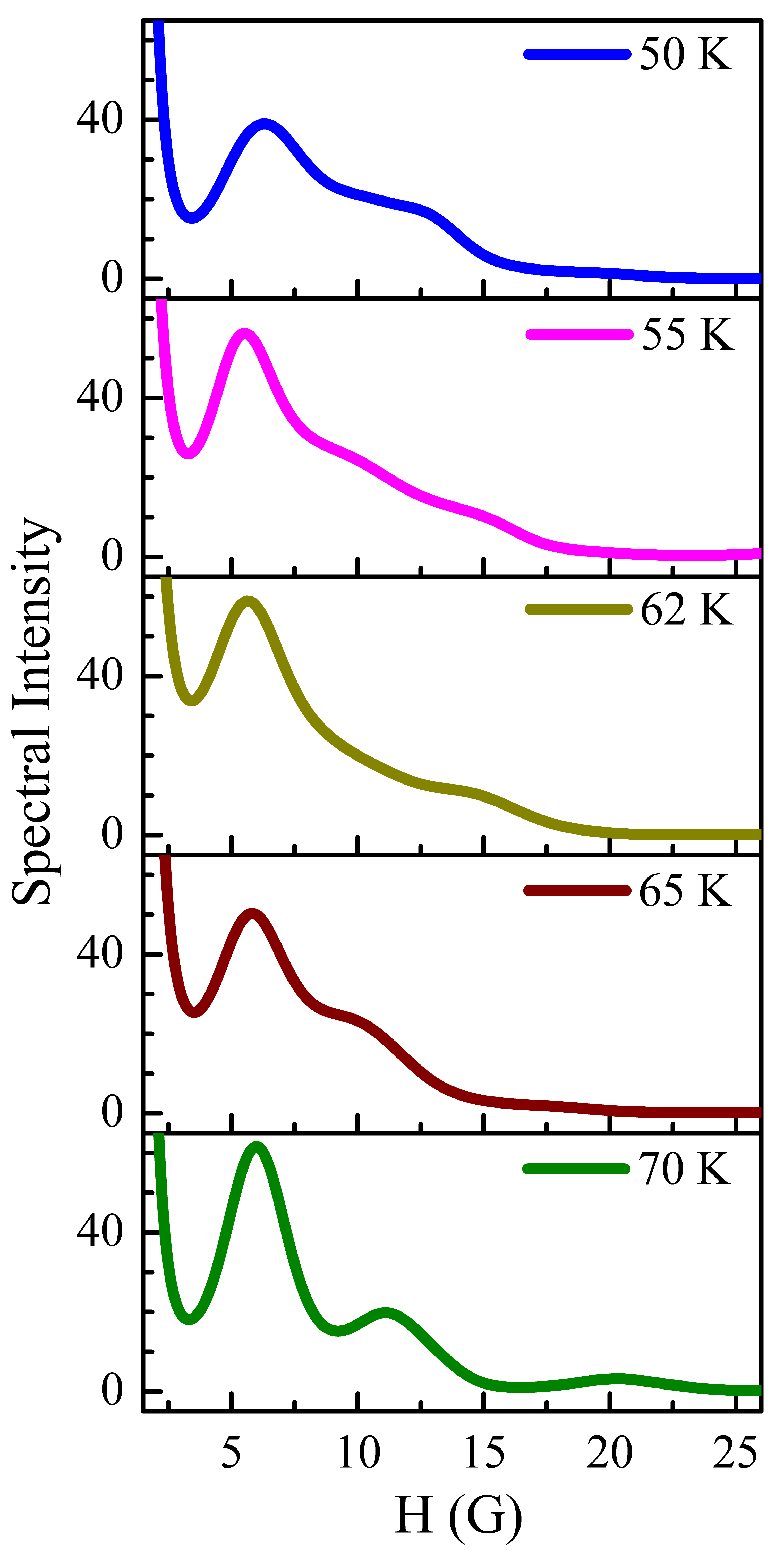}
    \caption{ Spectral intensity as a function of internal magnetic field in the temperature range of 50 K to 70 K, analyzed using the Maximum Entropy Method. The data illustrates the evolution of the internal magnetic field with temperature, providing insight into the underlying field distribution.}
    \label{Fig.9}
}
\end{figure}
 In Fig. 8, we have summarized the temperature dependence of the fitting parameters, which were obtained from ZF-$\mu$SR spectra of different temperatures. As shown in Fig. 8(a), a significant drop in the initial asymmetry is present around T$_{N}$ = 60 K, which signifies the long-range magnetic ordering (LRO) in this compound. The initial asymmetry drops by 0.42 around 60 K \cite{ref42,ref43}. However, we are not getting the 2/3 drop expected for the long range orderded magnetic materials. This discrepancy in the initial asymmetry drop is possibly due to the presence of short-range ordering (SRO) above the LRO along with a broad internal field distributions and significant magnetic damping also cause difficulties in tracing the exact initial asymmetry drop using ISIS pulsed muon source. Similar behavior is observed in many other well-known magnetic compounds at LRO, where it's found that the initial asymmetry drop at LRO is not exactly 2/3 due to the above-mentioned reasons \cite{ref40,ref41,ref42,ref43}. In $A_{2}$, there is a negligible amount of deviation as the value of $A_{1}$ is very high compared to $A_{2}$. In the paramagnetic region, the muon spin relaxation rate, $\lambda_{1}$ shows a nearly temperature-independent nature, which indicates that the relaxation is likely due to the exchange fluctuations of the Ru$^{4+}$ and the two Ru$^{5+}$ spins in the Ru-trimer (Fig. 8b). Using $k_{B} \Theta_{CW} = 2zS(S+1)J/3$ with z = 6 for both Ru$^{4+/5+}$ ions octahedra, S = 3/2 and 1 for Ru$^{5+}$ and Ru$^{4+}$, respectively, and $\Theta_{CW}$ = -118 K \cite{ref4}, we obtain the exchange fluctuation rate $\nu = \sqrt{z}JS/\hbar$ $\approx$ $6.02 \times 10^{11}$ and $7.5 \times 10^{11}$ $s^{-1}$ respectively for Ru$^{5+}$ and Ru$^{4+}$. So, using the relation $\lambda=2\Delta^{2}/\nu$ in the narrowing limit, $\lambda_{1}$ (T= 70 K) = 0.08262 $\mu s^{-1}$ gives a field distribution width $\Delta/\gamma_{\mu}$ $\approx$ 0.1851 T and 0.2069 T for Ru$^{5+}$ and Ru$^{4+}$, respectively. In the case of $\lambda_{2}$, possibly it's showing a temperature dependency in the paramagnetic region. However, due to the limited dataset, no definitive conclusions can be drawn from this.

Ba$_{5}$Ru$_{3}$O$_{12}$ is a trimer system, which contains disconnected Ru-trimers, in which two Ru$^{5+}$ and one Ru$^{4+}$ ions are present at three different Ru octahedra. In one of our previous papers, \cite{ref4}, we have shown the three Ru-octahedra in the trimer are distorted, and they have different ordered moments as 1.52, 1.36, and 0.91 $\mu_{B}$ for Ru1, Ru2, and Ru3. In oxide materials, the positive muons generally sit near the apical O$^{2-}$ in the Ru-octahedra \cite{ref34}. It was also predicted that, this system ordered antiferromagnetically with a canted spin state \cite{ref4}. For this antiferromagnetically ordered canted spin state, the spin moments are not perfectly antiparallel, leading to the development of a weak net magnetization in the ordered magnetic phase. This results in a broad distribution of local magnetic fields at the muon sites. As a result, we typically observe a gradual drop rather than a sharp drop in the initial asymmetry at the T$_{N}$ due to the non-uniform local field distributions. In our compound, we observe this same gradual drop in the initial asymmetry ($A_{1}$) at T$_{N}$, which strongly supports the canted (non-collinear) spin structure. The non-uniform local magnetic field is governed by different Ru-ground states predominantly arising from different degrees of RuO$_{6}$-octahedral crystallographic distortion.

We performed a maximum entropy analysis of the ZF-$\mu$SR data, as shown in Fig. 9. The results reveal two distinct peaks at all temperatures except 70 K, where three peaks appear. Examining the first peak, we observe that below the magnetic ordering temperature (T$_{N}$), the internal field increases due to the growth of the static ordered magnetic moment, confirming the presence of long-range magnetic order. At 70 K, the first peak broadens significantly, while the second peak becomes more pronounced, indicating enhanced spin dynamics. This suggests a weakening of static magnetism and stronger dynamic fluctuations, which is consistent with the presence of short-range ordering above T$_{N}$. Interestingly, at 70 K, a new peak emerges at a higher internal field ($\approx$ 21 G), which may arise due to various possible reasons, such as short-range magnetic ordering, polarized paramagnetic moments, or residual clusters of ordered magnetic moments above magnetic ordering. We attribute this feature to short-range magnetic ordering which agrees with the presence of INS excitation above ordering, large negative Curie-Weiss temperature of -118 K, and gradual drop in the initial asymmetry of ZF $\mu$SR around magnetic phase transition.

\section{Conclusions}
Here, we have documented the spin-wave excitation of the Ru-trimer system, Ba$_{5}$Ru$_{3}$O$_{12}$ through INS. The SpinW simulation replicates the experimental spin-structure and spin-excitations, revealing the various competing magnetic exchange interactions that play a decisive role in the magnetism of this trimer Ruthenate. Our results suggest a strong electronic correlation of Ru and possible spin-phonon coupling. The presence of INS spectra far above magnetic ordering manifests short-range correlation arising from isolated Ru-trimer. The $\mu$SR investigation demonstrates the non-uniform local magnetic fields arising from different Ru-atoms within the Ru-trimer. The temperature dependency of the fast relaxation rate at the paramagnetic region and the results from the maximum entropy analysis confirm the presence of short-range magnetic correlation above T$_{N}$. Finally, we conclude octahedral distortion and the exchange frustration govern a unique ground state for each Ru within the Ru-trimer and yield a non-collinear spin-structure, unlike all other Ruthenates belonging to nearly the same family. A small perturbation could tune the local structure and hybridization and control the magnetic ground state of ruthenium.
\section{Acknowledgement}
The authors thank Prof. Swarup Kumar Panda, Department of Physics, Bennett University, India, for the fruitful discussion.

T.B. greatly acknowledge the Science and Engineering Research Board (SERB) (Project No.: SRG/2022/000044), and UGC-DAE Consortium for Scientific Research (CSR) (Project No CRS/2021-22/03/544), Government of India for funding, and SEED Grant from RGIPT. The author gratefully acknowledges the use of resources at the Spallation Neutron Source (Inelastic Neutron Scattering), operated by Oak Ridge National Laboratory, USA, which contributed significantly to this research. The authors thank the ISIS Facility, STFC, UK, for muon beam time on the EMU instrument and the Department of Science and Technology, India, for financial support during the experiment. X.K. acknowledges the financial support by the U.S. Department of Energy, Office of Science, Office of Basic Energy Sciences, Materials Sciences and Engineering Division under Grant No. DE-SC0019259. D.T.A. thanks EPSRC UK for the funding (Grant No. EP/W00562X/1).J.S. would like to thank SERB, DST-India, for the Ramanujan Fellowship (Grant No. RJF/2019/000046).  M.K. thanks the University Grant Commission (UGC), India, for the research fellowship.
\bibliographystyle{apsrev4-2}

\bibliography{BaRuO}

\section{Supplementary}
S.I. Fig. 1 presents the Time-of-Flight neutron data, highlighting that the (010) magnetic Bragg peak is not reproduced when the magnetic moment of the Ru1 site is set to zero while refining the magnetic mnoment. This suggests that the (010) peak originates from the magnetic contribution of Ru1.
\renewcommand{\figurename}{S.I.Fig.}
\begin{figure}[h]
    \centering
    \includegraphics[width=0.5\textwidth,height=4in]{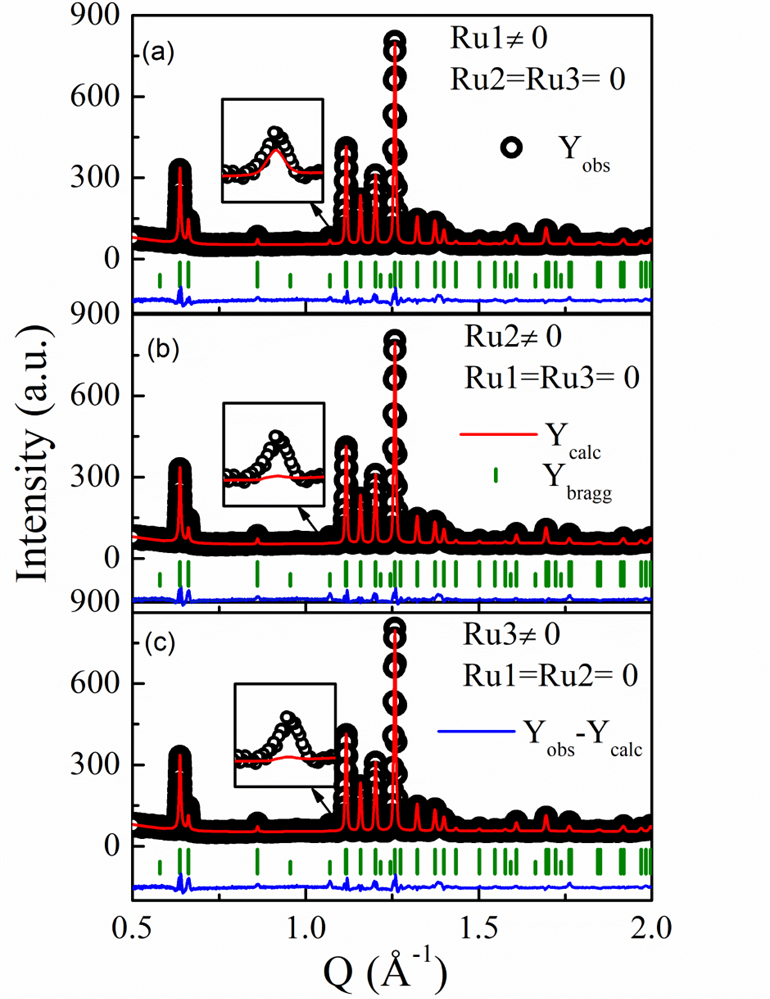}
    \caption{ Time-of-flight neutron data at 10 K (a) Ru2 = Ru3 = 0, (b) Ru1 = Ru3 = 0 and (c) Ru1 = Ru2 = 0.}
    \label{fig:example}
\end{figure}
\section{Exchange-interaction and ground state calculations}
The spin-Hamiltonian for this system is expressed as:
\begin{equation}\label{eq1}
\ H  = \sum_{i<j} \vec{S}_i^{\,\mathrm{T}} \mathbf{J}_{ij} \vec{S}_j + \sum_i \vec{S}_i^{\,\mathrm{T}} \mathbf{D} \vec{S}_i
\end{equation}
where  Jij is the 3 × 3 anisotropic exchange
tensor matrix given below:
\[
\mathbf{J}_{1} = 
\begin{pmatrix}
0 & 0 & 0 \\
0 & 0 & 0 \\
0 & 0 & 4.2
\end{pmatrix}
\quad \text{and} \quad 
\mathbf{J}_{2} = 
\begin{pmatrix}
0 & 0 & 1.8 \\
0 & 0 & 0 \\
0 & 0 & 0
\end{pmatrix}
\]
\[
\mathbf{J}_{3} = 
\begin{pmatrix}
0 & 0 & 0 \\
0 & 0 & 0 \\
4.5 & 0 & 0
\end{pmatrix}
\quad \text{and} \quad 
\mathbf{J}_{4} = 
\begin{pmatrix}
0 & 0 & 0 \\
0 & 0 & 0 \\
0.3 & 0 & 0
\end{pmatrix}
\]
\[
\mathbf{J}_{3} = 
\begin{pmatrix}
0 & 0 & 0 \\
0 & 0 & 0 \\
0.1 & 0 & 0
\end{pmatrix}
\quad \text{and} \quad 
\mathbf{J}_{4} = 
\begin{pmatrix}
0 & 0 & 0 \\
0 & 0 & 0 \\
-0.06 & 0 & 0
\end{pmatrix}
\]
\subsection{Trimer model for R\MakeLowercase{u}$_3$O$_{12}$}
We propose a model for the ground state and excited states of an isolated trimer system considering three intra-trimer exchange interactions. The Hamiltonian for the system is given by:

\begin{equation}
H = J_1 \mathbf{S_1} \cdot \mathbf{S_2} + J_2 \mathbf{S_1} \cdot \mathbf{S_3} + J_3 \mathbf{S_2} \cdot \mathbf{S_3}
\end{equation}

where \(J_1\) and \(J_2\) represent the nearest-neighbor exchange interactions, and \(J_3\) corresponds to the next-nearest neighbor exchange interaction. These interactions are defined for the Ru1-Ru2, Ru1-Ru3, and Ru2-Ru3 pairs, respectively. The spin operators \(\mathbf{S_1}, \mathbf{S_2}, \mathbf{S_3}\) represent the spin moments of Ru1, Ru2, and Ru3, where the spin moment is \(3/2\) for Ru1 and Ru2, and \(1\) for Ru3. Ru1 is positioned at the center of the trimer.

For complete characterization of the trimer states, additional quantum numbers \(S_{12}\) and \(S\) are required, arising from the vector sums:
\[
S_{12} = \mathbf{S_1} + \mathbf{S_2}, \quad S = \mathbf{S_1} + \mathbf{S_2} + \mathbf{S_3}
\]
with the constraints \(0 \leq S_{12} \leq 2S_1\) and \(|S_{12} - S_3| \leq S \leq |S_{12} + S_3|\), respectively. The trimer states are therefore defined by the wave functions \(|S_{12}, S\rangle\), and their degeneracy is \((2S + 1)\).

The eigenvalues corresponding to the trimer system are given by:
\begin{equation}\label{eq3}
\begin{split}
E(S_{12}, S) = & \frac{J_1}{2} \big[S_{12}(S_{12} + 1) - S_2(S_2 + 1) - S_1(S_1 + 1)\big] \\
               & + \frac{J_2}{2} \big[S(S + 1) - S_{12}(S_{12} + 1) - S_3(S_3 + 1)\big] \\
                & + \frac{J_3}{2} \big[S(S + 1) - S_{2}(S_{2} + 1) - S_3(S_3 + 1)\big]
\end{split}
\end{equation}

Putting all the values, we get:
\begin{equation}\label{eq3}
E(0, 1) = -\frac{15J_1}{4} - \frac{15J_3}{8}
\end{equation}
\begin{equation}\label{eq3}
E(1,0) = -\frac{11J_1}{4} - 2J_2 - \frac{23J_3}{8}
\end{equation}
\begin{equation}\label{eq3}
E(1,1) = -\frac{11J_1}{4} + J_2 + \frac{J_3}{8}
\end{equation}
\begin{equation}\label{eq3}
E(1,2) = -11J_1 + J_2 + \frac{J_3}{8}
\end{equation}
\begin{equation}\label{eq3}
E(2,1) = -\frac{7J_1}{4} - 2J_2 - \frac{23J_3}{8}
\end{equation}
\begin{equation}\label{eq3}
E(2,2) = -3J_1 - J_2 + \frac{J_3}{8}
\end{equation}

\begin{equation}\label{eq3}
E(2,3) = -\frac{7J_1}{4} + J_2 + \frac{J_3}{8}
\end{equation}
\begin{equation}\label{eq3}
E(3,2) = -\frac{3J_1}{4} - 2J_2 - \frac{23J_3}{8}
\end{equation}
\begin{equation}\label{eq3}
E(3,3) = \frac{9J_1}{4} - J_2 +\frac{25J_3}{8}
\end{equation}
\begin{equation}\label{eq3}
E(3,4) = -\frac{3J_1}{4} + J_2 + \frac{J_3}{8}
\end{equation}
\\
From INS data, we get the two excited states: first \(\varepsilon_1 = 5.6 \, \text{meV}\) and second \(\varepsilon_2 = 10 - 15 \, \text{meV}\). By Gaussian fitting of that data, we get one peak around 5.6 meV excitation  and two peaks in between  \(10 - 15 \, \text{meV}\) excitations.
From simulation of spin wave using SpinW software We get  \(J_1 = 4.2 \ \text{meV}\) \(J_2 = 1.8 \ \text{meV}\) and \(J_3 = 4.5 \ \text{meV}\), by putting these values in above equation, we get the following excited states:\\

\textbf{First Excited State:} \\

Transition from \(|3,2\rangle \rightarrow |2,2\rangle\)
\[
\varepsilon_1 = E(2,2) - E(3,2)
\]
\[
\varepsilon_1 = \frac{7J_1}{4} + 2J_2 + \frac{23J_3}{8} - \left[ \frac{3J_1}{4} + 2J_2 + \frac{23J_3}{8} \right]
\]
\[
\varepsilon_1 =  5.85 \, \text{meV}
\]

\textbf{Second Excited State:} \\

Transition from \(|3,2\rangle \rightarrow |1,1\rangle\)
\[
\varepsilon_2 = E(1,1) - E(3,2)
\]\\
\[
\varepsilon_2 = 3J_1 + J_2 - \frac{J_3}{8} - \left[ \frac{3J_1}{4} + 2J_2 + \frac{23J_3}{8} \right]
\]
\[
\varepsilon_2 = 10.5 \, \text{meV}
\]

\textbf{Third Excited State:} \\

Transition from \(|3,2\rangle \rightarrow |2,3\rangle\)
\[
\varepsilon_3 = E(2,3) - E(3,2)
\]
\[
\varepsilon_3 = \frac{7J_1}{4} - J_2 - \frac{J_3}{8} - \left[ \frac{3J_1}{4} + 2J_2 + \frac{23J_3}{8} \right]
\]
\[
\varepsilon_3 = 14.7  \, \text{meV}
\]
\subsection{Origin of excitation at high-$|$Q$|$ region}
\begin{figure}[h]
    \centering
    \includegraphics[height=3.5in, width=3.5in]{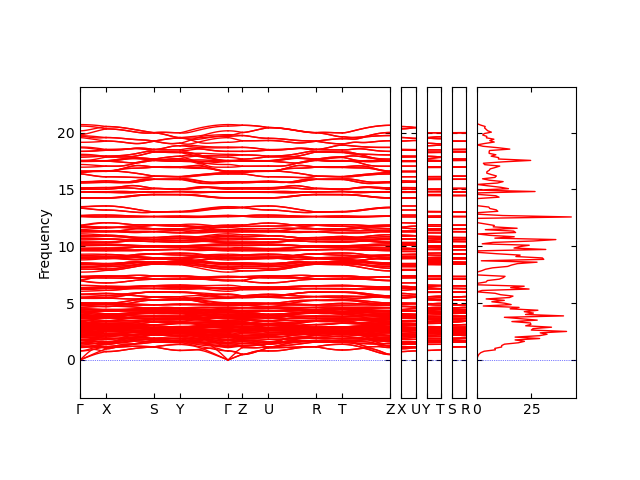}
    \caption{Calculated phonon modes for Ba$_5$Ru$_3$O$_{12}$ based upon pre-trained machine learning force fields as described in the main text. Vertical axis is the phonon energy in units of THz. (1 THz = 4.136 meV). Panels indicate wave-vector dependence in reciprocal space using standard reciprocal space coordinates for the orthorhombic crystal structure. The panel on the far right is the density of states (horizontal axis) as a function of energy transfer (vertical axis). }
    \label{fig:example}
\end{figure}
    \begin{figure}[h]
\centering
    \includegraphics[height=4in, width=4in]{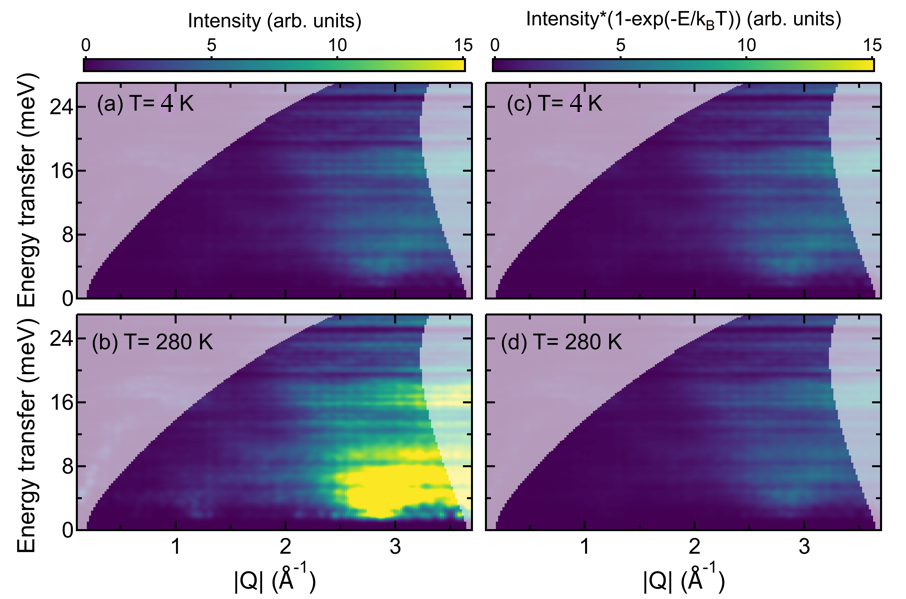}
    \caption{Calculated phonon scattering intensity (a)-(b) and Bose factor normalized scattering intensity (c)-(d) for Ba$_5$Ru$_3$O$_{12}$. Calculations are based upon the MLFF phonon calculations shown in Fig. S.I. 3. The neutron scattering intensity was convolved with the calculated instrumental energy resolution for the E$_i$ =30 meV measurements. The wave-vector resolution was determined from the width of the aluminum nuclear Bragg peaks measured in the empty can measurements.  The calculation was also convolved with this determined wave-vector resolution. All panels are shown in the same intensity scale. The lightly shaded regions are outside of the measured range of energy and wave-vector transfer allowed from the kinematic constraints and the detector coverage of the instrument. }
    \label{fig:example}
\end{figure}
\renewcommand{\figurename}{S.I.Fig.}
\begin{figure}[h]
    \centering
    \includegraphics[height=4.5in]{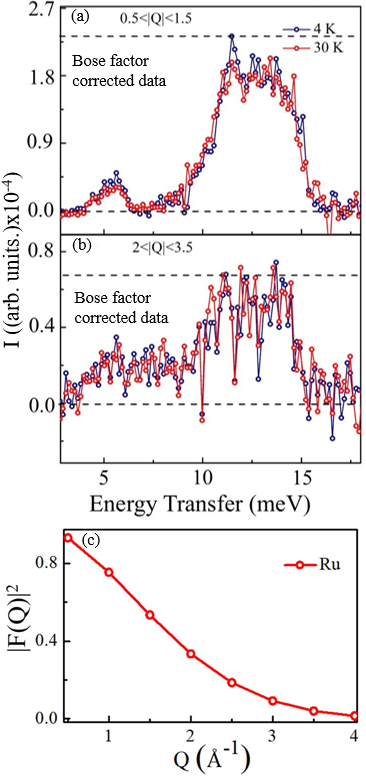}
    \caption{Bose factor corrected INS spectra (a) at low-$|$Q$|$ range (0.5$<$ $|$Q$|$$ <$ 1.5), showing strong magnetic excitation near 5-15 meV at 4 K and 30 K. The horizontal dashed line indicates the maximum intensity used for scaling to high-$|$Q$|$. (b)at high-$|$Q$|$ range (2$<$ $|$Q$|$$ <$ 3.5). The dashed line denotes the maximum magnetic intensity estimated by scaling the low-$|$Q$|$ value. (c) Squared magnetic form factor $|$F(Q)$|$$^2$ for Ru as a function of momentum transfer $|$Q$|$, calculated using standard analytical expressions for 4\textit{d}transition metals.}
    \label{fig:example}
\end{figure}

\end{document}